\documentclass[12pt]{article}
\usepackage{rotating}
\usepackage{indentfirst}
\usepackage{cite}
\usepackage{doublespace}
\usepackage{citesupernumber}
\usepackage{epsfig}
 \tiny \normalsize

\title{Optimal Allocation of Replicas to Processors
in Parallel Tempering Simulations}

\author{David J.\ Earl and Michael W.\ Deem\\
Department of Bioengineering and Department of Physics \& Astronomy\\
Rice University\\
6100 Main Street---MS 142\\
Houston, TX\ \ 77005--1892\\[0.1in]
}

\begin{document}
\maketitle
Corresponding author: M.\ W.\ Deem, mwdeem@rice.edu, fax: 713--348--5811.
\clearpage
\newpage
\begin{abstract}
The optimal allocation of replicas to a homogeneous or
heterogenous set of processors is derived
for parallel tempering simulations on multi-processor machines.
In the general case, it is possible
without substantially increasing wall clock time 
to achieve nearly perfect utilization of CPU time.
Random fluctuations in the execution times of each of the replicas do
not significantly degrade the performance of the scheduler.
\end{abstract}

\section{Introduction}
\label{sec:introd}
The parallel tempering, or replica-exchange, Monte Carlo method is
an effective molecular simulation technique for the study of
complex systems at low temperatures.\cite{geyer91,geyer95,marinari98}
Parallel tempering achieves good 
sampling by allowing systems to escape from low free energy minima by 
exchanging configurations with systems at higher temperatures, which are
free to sample representative volumes of phase space.   The
use of parallel tempering is now widespread in the scientific community.

The idea behind the parallel tempering
technique is to sample $n$ replica systems, each in the canonical 
ensemble, and each at a different temperature, $T_{i}$.  Generally
$T_{1} < T_{2} < ... < T_{n}$, where $T_{1}$ is the low temperature system,
of which we are interested in calculating the properties.  Swaps,
or exchanges,
of the configurational variables
between systems $i$ and $j$ are accepted
with the probability 
\begin{equation}
\label{eqn:acc}
p = \min\{1,~\exp\left[-(\beta_{i}-\beta_{j})(H_{j}-H_{i})\right]\},
\end{equation} 
where $\beta=1/(k_{\rm B} T)$ is the reciprocal temperature, and
$H_{i}$ is the Hamiltonian of the configuration in system $i$.
Swaps are typically attempted between systems with adjacent 
temperatures, $j=i+1$.  Parallel tempering is an exact method
in statistical mechanics, in that it satisfies the detailed balance
or balance condition,\cite{Deem_balance} depending on the implementation.

Due to the need to satisfy the balance condition,
the $n$ different systems must be synchronized 
whenever a swap is attempted.  This synchronization is
in Monte Carlo steps, rather than in real, wall clock time.
In other words, all processors must finish one Monte Carlo step
before any of the processors may start the next Monte Carlo step.
In parallel tempering, a convenient definition of
Monte Carlo step is the ordered
set of all of the Monte Carlo moves that
occur between each attempted swap move.
These Monte Carlo moves are all of the individual moves
that equilibrate each system in the parallel tempering ensemble, such as
Metropolis moves, configurational bias moves, volume change moves,
hybrid Monte Carlo moves, and so on.
Rephrasing, the balance condition requires that
at the beginning of each Monte Carlo step, each replica must have 
completed the same number of Monte Carlo steps and must be available 
to swap configurations with the other replicas.  This constraint
introduces a potentially 
large inefficiency in the simulation, as different replicas are
likely to require different amounts of computational processing 
time in order to complete a Monte Carlo step.  This inefficiency 
is not a problem on a single processor 
system, as a single processor will simply step through all the
replicas to complete the Monte Carlo steps of each.
This inefficiency is a significant problem on multi-processor machines,
however, where 
individual CPUs can spend large quantities of time idling as they wait
for other CPUs to complete the Monte Carlo steps of other replicas.

Traditionally, each processor on multi-processors machines has been
assigned one replica in parallel tempering simulations.
It is the experience of the authors that this type of
assignment  is generally
highly inefficient, with typical CPU idle times of
40--60\%.   When one takes into account that the lower-temperature systems
should have more moves per Monte Carlo step due to the increased correlation
times, the idle time rises to intolerable levels that can approach 95\%.
The issue of idle time has not been previously addressed, and it is clear
that a scheme which could allocate replicas to processors in an optimal 
manner would be useful.

In this paper we address the optimal allocation of replicas to CPUs in 
parallel tempering simulations.  The manuscript is organized as follows.  
In Sec.~\ref{sec:theory} we present the theory for the allocation of 
replicas to a homogeneous set of
processors.  In Sec.~\ref{sec:results} we present results 
where the theory is applied to several model examples.  
In Sec.~\ref{sec:discussion} we discuss our results, compare 
them with the conventional parallel tempering 
scheme, and consider the effects of including communication times and 
randomness in execution time into our model.  We draw our conclusions in 
Sec.~\ref{sec:conc}.  An appendix presents the theory for a
heterogeneous set of processors.

\section{Theory of Replica Allocation to Processors}
\label{sec:theory}

In a parallel tempering simulation, balance requires that 
each replica system be synchronized at the start of each
Monte Carlo step.
Considering replica $i$, in every Monte Carlo
step we will attempt $N_{\rm move}(T_{i})$ random Monte Carlo configurational
moves, and the average real 
wall clock time to complete one Monte Carlo move is given by 
$\alpha(T_{i})$.  The total wall clock time for replica $i$ to 
complete its Monte Carlo step is
\begin{equation}
\label{eqn:tau}
\tau_{i}=\alpha(T_{i})N_{\rm move}(T_{i}).
\end{equation}
As we have already stated, the simple allocation of one replica 
to one processor for the entire simulation is inefficient.  This 
is because $\alpha$, the time per configurational move, depends on the
temperature of the system. The value of $\alpha$ can typically vary by a
factor of 3 or more between the fastest and the 
slowest system resulting in long idle times 
for the CPUs that are assigned to the higher temperature systems.  
The value of $\alpha$ varies because the composition of the 
configurational moves and their acceptance ratio varies with
temperature.  Typically, but not always, the highest temperature moves
take less wall clock time on average to complete.
Additionally, it is often desirable to perform more configurational
Monte Carlo moves per 
Monte Carlo step at lower temperatures because the correlation time is 
longer than it is at higher temperatures.  This makes the inefficiency 
of allocating one replica to one processor
dramatically worse.  In Eqn.~\ref{eqn:tau}, $N_{\rm move}$ 
is a function of $T_{i}$ to allow for the larger number of
configurational moves that may be performed at lower
temperatures.  In most simulations that are 
currently performed, $N_{\rm move}$ is the same for all replicas 
because of the disastrous inefficiency implications of increasing 
$N_{\rm move}$ for low temperature replicas, for which $\alpha$ is also
often larger.  
Using an optimal allocation of replicas, the possibility of 
varying $N_{\rm move}$ for different replicas exists, as 
discussed in Sec.~\ref{sec:results} below.

The optimal allocation of replicas to processors  is a non-trivial
problem even in remarkably simple situations.  For  example, consider
the case where $n=3$, $\tau_{1}=5$, $\tau_{2}=4$, and $\tau_{3}=3$.
Using three processors is clearly inefficient, as two processors would be 
idle while they are waiting for replica 3 to complete.  
The optimal allocation is to split one of the replicas on two
processors, as shown in Fig.\ \ref{fig:basic}. Only two 
processors are required, and they will both run at 100\% efficiency if the 
replica is split correctly.  Note that the splitting must be causally
ordered in time.
In the example of Fig.\ \ref{fig:basic}, replica 2 is started on
processor 2 and completed on processor 1 two time units after being
stopped on processor 2.

A general replica scheduler can be derived
starting with the assumptions that one replica cannot be 
simultaneously run on more than one processor
and that one processor can only run one replica at a time, this second
assumption being the simplest and, as it turns out, the most efficient
use of the processors.
The logic  of the derivation
comes from scheduling theory,\cite{Coffman,Ashour} which is frequently used 
to solve problems of this type in operations research and industrial 
engineering.  Given $n$ replicas, where the time to complete 
replica $i$ is $\tau_{i}$, the total processing time required to complete 
all of the replicas is
\begin{equation}
\label{eqn:tot_time}
W=\sum_{i=1}^{n} \tau_{i}.
\end{equation}
We let $\tau_{\rm long}$ be the CPU time of the longest replica.  
If we have $X$ processors, then the shortest possible total wall clock time
required to 
complete execution of all of the replicas is given by
\begin{equation}
\label{eqn:time_eqn}
\tau_{\rm wall}=\max(W/X, \tau_{\rm long}).
\end{equation}
The optimum integer number of processors to achieve 100\% theoretical 
efficiency will be
\begin{equation}
\label{eqn:X_n}
X^{(N)}=\lfloor W/\tau_{\rm long} \rfloor,
\end{equation}
where $\lfloor y \rfloor$ is the largest integer equal to or less than
the real number $y$.
The number of processors required to achieve the minimum wall clock time 
will be
\begin{equation}
\label{eqn:X_n+1}
X^{(N+1)}=\lceil W/\tau_{\rm long} \rceil ,
\end{equation}
where $\lceil y \rceil$ is the smallest integer equal to or greater
than the real number $y$.
The optimal allocation can either be done for minimum, zero percent,
idle time, $X^{(N)}$, or minimum wall clock time
$X^{(N+1)}$.
Having made the choice of one of these two numbers of processors,
the optimal scheduler  then proceeds by assigning 
the replicas 
sequentially to the first processor until 
that processor has filled its allocation of $\tau_{\rm wall}$ wall
clock time.  
Typically this will result in the last replica allocated to the first processor
being split, with the ``remaining'' time carried over to the second processor.  
The remaining replicas are sequentially allocated to the second processor,
with again a possible split in the last replica allocated.  This procedure
is repeated until all the replicas have been allocated.
In the minimum wall clock, $X^{(N+1)}$, case, the final processor will
not be completely filled unless $W/X^{(N+1)}=\tau_{\rm long}$, and there
will be a small amount of idle time.  In the minimum idle time case,
there will be no idle time.
  An example of how the 
scheduler assigns replicas to processors is shown in Figure~\ref{fig:sched} 
for a 20 replica case where $\tau_{\rm long} / \tau_{\rm short} = 3$, where
$\tau_{\rm short}$ is the wall clock time of the replica that
completes its Monte Carlo step most quickly.

It is immediately apparent that the scheduler\cite{scheduler} is 
extremely simple and very effective. The scheduler may easily 
be applied to existing parallel simulation codes.   To apply the
theory to a practical simulation, one must first 
perform a short preliminary 
simulation for each replica to obtain an estimate of $\alpha(T_i)$,
and hence $\tau_{i}$ from Eqn.~\ref{eqn:tau}.  We note that the 
scheduler could be run after each Monte Carlo step, since
the calculation time associated with the scheduler is minimal.
Such use of the scheduler would automatically lead to an adaptive
or cumulative estimate of $\alpha$.  Note that at all times, the
balance properties of the underlying Monte Carlo scheme are unaffected by
the replica allocations of the scheduler.  It 
is also worthy of comment that the scheduler could be run with parallel 
tempering in multiple dimensions, for example differing chemical potentials
\cite{Yan99,Yan00,Faller02} or
pair potentials\cite{Bunker01} for each replica,
in addition to variations in temperature. Increasing the number 
of order parameters that we use in the parallel tempering not only may
improve sampling but also may provide a better estimate of $\alpha$,
since the estimate of $\alpha$ as a local function of phase space
increases as the number of order parameters increases.

In this section we have derived the scheduler for a homogeneous
cluster of processors.  In the Appendix we derive a similar scheme for a
heterogeneous cluster.

\section{Results}
\label{sec:results}
In this section we apply the optimal replica
scheduler to three different parallel tempering simulation 
examples.  Details of the three different examples are 
given below, and the performance of the scheduler
can be seen in Table \ref{table1}.  
Results are shown in the table for the minimum idle time,
minimum wall clock time, and traditional 
one-replica-per-processor cases.  
For each case we show the number of processors used, the 
CPU idle time as a percentage of the overall time for one 
Monte Carlo step, and the real wall clock time for the simulation 
relative to that of the traditional parallel tempering 
approach.  To motivate the parameter values chosen for the
examples, we note than in our experience with simulations of
the 20--50 amino acid peptides from the innate immune system
that are known as cystine-knot peptides, 
we find the ratio of correlation times between the low and
high temperature replicas can vary by a factor of
$10^2$ to $10^5$,
$N_{\rm move}(T_{1})/N_{\rm move}(T_{n})=10^2$--$10^{5}$, 
on the order of 
$N_{\rm move}(T_{n})=10^{3}$--$10^{5}$ configurational Monte Carlo
moves are typically performed during each Monte Carlo step at the
highest temperature, and
$N_{\rm move} =  10^{6}$ configurational Monte Carlo moves 
take on the order of 24 hours to complete.

\subsubsection*{Example 1}
For example 1, the simulation system is chosen such that 
$n=20$, and $\alpha(T_{1})/\alpha(T_{n})=3$.  In parallel 
tempering simulations, it is usual for the temperature to 
increase exponentially from $T_{1}$ to $T_{n}$,
since higher temperature systems have wider energy histograms, and so 
higher temperature replicas can be spaced more widely than lower 
temperature replicas.\cite{Kofke2002}
For specificity, we assume that the wall clock time per configurational
step also increases exponentially from
$\alpha(T_{n})$ to $\alpha(T_{1})$.  
We take $N_{\rm move}$ to be 
constant for each of the replicas.  The allocation 
of the replicas to the different processors is shown in 
Figure~\ref{fig:sched}a) and b) for the traditional and
zero idle time cases, respectively.  
This example is typical of most parallel tempering simulations 
that are currently being performed on multi-processor systems.

\subsubsection*{Example 2}
For example 2, we use $n$ and $\alpha(T_{i})$ from example 1.  
We, furthermore,  consider that the correlation times of the lower 
temperature replicas are longer, and so there should be more
configurational moves per Monte Carlo step at the lower temperatures.
We consider $N_{\rm move}$ to increase exponentially from 
$N_{\rm move}(T_{n})$ to $N_{\rm move}(T_{1})$ such that 
$N_{\rm move}(T_{1})/N_{\rm move}(T_{n})=100$.  With the 
values for $\alpha(T)$ from example 1, we find
$\tau_{\rm long}/\tau_{\rm short}=300$.

\subsubsection*{Example 3}
For example 3, we use $n=50$, modeling $\alpha(T_{i})$ in the 
same way as in examples 1 and 2, $\alpha(T_{1})/\alpha(T_{n})=3$.  
We model $N_{\rm move}$ in the same way as in example B, but in 
this example set $N_{\rm move}(T_{1})/N_{\rm move}(T_{n})=1000$,
since the reason for the increased number of replicas would have
been the poor and slow equilibration at the lowest temperatures.
We find $\tau_{\rm long}/\tau_{\rm short}=3000$.

\section{Discussion}
\label{sec:discussion}
From Table \ref{table1}, it is clear that the scheduler 
substantially improves the CPU utilization in parallel 
tempering simulations.  This allows the multi-processor
cluster to be used with confidence, for example, for
other jobs or simulations at other parameter values
of interest.  Example 1 demonstrates that 
the number of processors used can be reduced by 40\%
with an increase of only 1.66\% in wall 
clock time.  Alternatively, the number of processors 
can be reduced by 35\% and result in no increase in wall clock time
relative to the
traditional parallel tempering method.  As Example 1 is conservative
in its characterization of most
multi-processor parallel tempering simulations currently being
performed, we anticipate that utilization of the 
optimal scheduler presented here will result in a large 
increase in the computational efficiency of parallel tempering simulations.

It is interesting to note that, for all examples, 
as we increase the number of 
processors used in the simulations, $X$, from 1, the wall clock 
time decreases until the number of processors 
that result in minimum wall clock time is used,
$X^{(n+1)}=\lceil W/\tau_{\rm long} \rceil$. Increasing the 
number of processors still further, to say the number of replicas,
results in no reduction in 
overall simulation time and only increases the CPU idle time.  
This behavior is
demonstrated in Figure~\ref{fig:idle}, where the idle time is shown 
as a function of $X$ for example 2.  This figure highlights the importance of
proper job scheduling on large, multi-processor clusters.
The use of the optimal scheduler 
derived here is needed in order for the simulation to make the 
best use of a large number of CPU cycles.  It is theoretically possible to 
achieve 100\% 
efficiency on multi-processor systems, making them 
ideal for parallel tempering simulations.  This is especially 
important in cases where it is desirable to vary $N_{\rm move}$ 
between different replicas (examples 2 and 3).  Taking into account the
dependence of the correlation time on temperature is
computationally disastrous for the traditional one-replica-per-processor 
method of performing parallel tempering simulations, as CPU idle times 
easily become $>$ 90\%.  
However, the optimal scheduler makes the simulation of 
this case feasible, opening 
the door to performing parallel tempering
simulations that sample configurational 
space more effectively and efficiently.

In the results presented in Sec.~\ref{sec:results}, we have not
explicitly taken into account communication times or the time taken 
to conduct swap moves.  Swap moves that exchange 
configurations between replicas 
occur at the beginning of each Monte Carlo 
step and replica allocations occur at the beginning and possibly once
within each Monte Carlo step.  These
operations are extremely rapid compared to the $N_{\rm move}$ configuration
moves  performed for each replica, as one can show.
Recalling from the
Results section that one configurational move takes 
approximately 0.1 seconds and knowing that a typical communication time 
for inter-processor message passing is
on the order of $10^{-4}$ seconds, we find that example 3 contains the 
most communication time.  In example 3, the
increase in idle time due to communication from the zero idle time 
case is less than 0.00001\%.  This demonstrates that communication 
time is not a significant effect in these types of simulations.
Communication effects can, thus, safely be ignored.

We have characterized the execution time of each replica in a
deterministic fashion, but in reality the execution time is a
stochastic quantity due to noise in variables not among the
degrees of freedom chosen for the parallel tempering.
In order to model the simulation times more realistically, we 
have also included randomness into our analysis.  That is,
the value of $\alpha$ is assumed to fluctuate during each
configurational step.  As previously mentioned, the accuracy of the 
estimation of $\alpha$ is dependent on the number of order 
parameters used to parameterize it.  Thus, fluctuations in $\alpha$ 
will be smaller for systems that use parallel tempering in multiple 
dimensions.  We note that for the case where the temperature is 
the only parameter used to characterize $\alpha$, fluctuations in 
$\alpha$ can be as high as 10-50\%.  This results in a fluctuation in 
the time required to
complete replica $i$, which can be represented mathematically as
\begin{equation}
\label{eqn:time_random1}
\tau_{i}=\alpha(T_{i}) N_{\rm move}(T_{i})
 \left\{
1+\frac{\sigma}{ [N_{\rm move}(T_{i}) /
                  \beta(T_{i})]^{1/2} }
\right\},
\end{equation}
where $\sigma$ is a Gaussian random number, and $\beta$ is a value that is 
proportional to the correlation time.  As we generally choose $N_{\rm move}$ 
to be proportional to the correlation time, we expect 
$N_{\rm move}/\beta=$ constant.  Thus, we use
\begin{equation}
\label{eqn:time_random2}
\tau_{i}=\alpha(T_{i}) N_{\rm move}(T_{i}) [1+\gamma \sigma ]
\end{equation}
to model the fluctuations.  We examine the cases where 
$\gamma=0.1, 0.5$, and $1.0$.  To analyze the performance of the
scheduler in the presence of the randomness,
we take into account that 
a processor may be idle while it is 
waiting for another processor to complete its share of 
calculations on a replica system that is shared between the 
two processors.

Table \ref{table1} shows the results of including randomness into 
our model for examples 1--3.  The averages and standard errors 
are calculated from the average results from 10 blocks, each containing 1000 
runs of the simulation system.  The CPU idle time 
increases monotonically and non-linearly with $\gamma$.
 For the 
more complex systems where $N_{\rm move}$ is varied, the
inefficiency introduced by the randomness is smaller, since
the randomness of several replicas is typically averaged over on
most of the processors.
The results are encouraging and 
show that the efficiency of the parallel tempering
simulations organized by the 
scheduler remains within an acceptable limit, even when relatively 
large fluctuations are considered.  Increasing 
$N_{\rm move}$ will lead to lower fluctuations, with the
observed efficiency converging to the $\gamma \to 0$ limit as
$O(1/N_{\rm move}^{1/2})$.

\section{Conclusions}
\label{sec:conc}
In this paper we have introduced a theory for the optimal allocation 
of replicas to processors in parallel tempering simulations.
The scheduler leaves intact the balance or detailed balance properties
of the underlying parallel tempering scheme.  The 
optimal scheduler derived from the theory
allows multi-processor machines to be efficiently used for
parallel tempering simulations.  The allocation of replicas to CPUs
produced by the scheduler results in a significant enhancement of
CPU usage in comparison to the traditional one-replica-per-processor
approach to multi-processor parallel tempering.
The optimal scheduling vastly reduces the number of required 
processors to complete the simulation, allowing an increased number of
jobs to be run on the cluster.  The computational efficiency of the
scheduler also makes it feasible to vary the number
of configurational moves per Monte Carlo step, which was not
practicable using the one-replica-per-processor scheme, due 
to the associated large inefficiencies.
This flexibility to vary number of configurational steps is desirable
because the correlation time at 
lower temperatures is often much longer than that at higher temperatures.  

Our results show that randomness does not have a significant effect for 
$\gamma<0.1$, and the performance is still quite tolerable even for
the extreme case of $\gamma = 1$.  Despite the random execution times,
the replica allocation produced by the optimal
scheduler is always significantly more efficient than the
traditional one-replica-per-processor approach.
The idle time caused by random execution times is reduced as the
number of configurational moves per Monte Carlo step is increased.
Furthermore, parallel tempering in more than one 
dimension, with order parameters other than temperature, allows for
a more accurate determination 
of the CPU time per replica.
For the same reason, these extra dimensions will also aid 
the sampling efficiency of the underlying parallel tempering algorithm.

\section*{Acknowledgments}
This work was supported by the U.S.\ Department of Energy Office of Basic
Energy Sciences.

\renewcommand{\theequation}{A-\arabic{equation}}
\setcounter{equation}{0}

\section*{Appendix}
\subsubsection*{Allocation Scheme for a Heterogeneous Cluster}
Using scheduling theory\cite{Coffman,Ashour} it is possible to 
derive an allocation scheme for a multi-processor machine with
heterogeneous processors.  It is assumed that
the number of CPU cycles required for each replica to complete one
Monte Carlo step and the speed of each of the processors in the machine
are known.  In this general scheme, the number of processors used
by the scheduler, $m$, is adjusted downward until
an acceptably low idle time and total
wall clock time are achieved.

For $n$ replicas, where  $\tau_{i}$ is
the number of CPU cycles required to complete
replica $i$, the total number of CPU cycles required,
$W$, is given in Eqn.~\ref{eqn:tot_time}.  We now define
\begin{equation}
\label{eqn:A1}
W_{j}=\sum_{i=1}^{j} \tau_{i}, 1\leq j\leq n.
\end{equation}
For $m$ processors, where $k_i$ is the speed of
each processor in CPU cycles per unit time, with
 $k_{1}\geq k_{2}\geq ...\geq k_{m}$, 
the total number of CPU cycles available per unit time is
\begin{equation}
\label{eqn:A2}
K=\sum_{i=1}^{m} k_{i}.
\end{equation}
We define
\begin{equation}
\label{eqn:A3}
K_{j}=\sum_{i=1}^{j} k_{i}, 1 \leq j \leq m.
\end{equation}
The shortest possible wall clock time
to execute the Monte Carlo step for all the replicas is then
\begin{equation}
\label{eqn:A4}
\tau_{\rm wall}=\max(W/K, \tau_{\rm long}),
\end{equation}
where $\tau_{\rm long}$ is
the maximum value of $W_{j}/K_{j}, 1 \leq j \leq m$.

The general scheduler works with a time interval granularity of $dt$.
At the start of the simulation and
at the end of each time interval, we assign a level of priority to
the replicas.  The highest priority is given to the replica with
the largest number of CPU cycles required for completion, and
the lowest priority is given to the replica with the least number of CPU
cycles remaining.  A loop is performed through the priority levels,
starting at the highest priority.  If there are $r$ replicas in the
priority level under consideration and $s$ remaining
 unassigned processors and if $r\leq s$, then 
the $r$ replicas are assigned to be executed on the
fastest $r$ number of processors.  If the processors have different
speeds, each replica must spend an equal amount of wall clock
time on each of the processors during the time interval, $dt$.
The total wall clock time for the
step is computed from the processor speeds and the required
number of CPU cycles.  The number of configurational moves
that equals $1/r$ of the wall clock time on each processor is computed, and
this number
is the number of configurational moves that each replica will perform on each
processor.  For the first $1/r$ of the wall clock time,
the replicas are assigned sequentially to the $r$ processors.
For the next $1/r$ of the wall clock time, the assignment of the
replicas to the processors is cyclically permuted, \emph{i.e.}\ replica 1 to
processor 2, replica 2 to processor 3, \ldots, replica $r$ to processor 1.
The assignment of replicas to processors is cycled at the end of
each $1/r$ of wall clock time until the entire time step is completed.
On the other hand
if $r > s$, the replicas are assigned to
the processors by splitting the time interval in each processor $r$
times, and assigning the replicas to spend one short time interval being
processed in each processor.  This is accomplished by assigning
the first processor to execute sequentially replicas $1, 2, \ldots, r$.
The second processor is assigned a cyclic permutation of the
replicas to execute sequentially: replicas $2, 3, \ldots, r, 1$.
In general processor $i$ executes a cyclic permutation of the
replica sequence of processor $i-1$.  This allocation leads to 
each replica being executed for an equal amount of wall clock time on
each processor.  A singe replica, moreover, is never allocated to
more than one processor at a single point in time.

If there are still processors remaining to be allocated, the
replicas at the next lower priority level are allocated by 
this same process.  The procedure is repeated until all processors
have been allocated or all replicas have been allocated.

The replica assignment for wall clock time $dt$ is now complete.
Replica are reassigned for the next period of wall clock time
using the same rules.  If the time interval,
$dt$, is chosen to be small enough, then the total wall clock
time of the simulation tends toward $\tau_{\rm wall}$.
After the wall clock time of the entire Monte Carlo step has been assigned,
the simulation can be performed.

There is some flexibility in the use of this general optimal scheduler for
a heterogeneous multi-processor machine.  In general, the best value
of $m$ is not known in closed form.  It is found by choosing the smallest
value of $m$ that gives an acceptably low value wall clock time,
Eqn.~\ref{eqn:A4}, and an acceptably low idle time in the
derived allocation.  The time step for the scheduler, $dt$, must also be
chosen.  It should be chosen to be small, but not so small that communication
effects become significant.  Moreover, 
there must be many configurational Monte Carlo
steps per time step, $dt$, otherwise the splitting of replicas among
$r$ processors required by the algorithm will not be possible.
The computational time associated with the scheduler will
generally be very much smaller than that associated with the simulation.
The scheduler may, therefore, be run after each Monte Carlo step.
Such use of the scheduler would automatically lead to an adaptive
or cumulative estimate of the execution times required by each replica.

In practical application of the results of this general scheduler, the
processor allocation will typically be reordered
to an equivalent one.  For example, in the case of two replicas of equal
length to be assigned to a single processor, 
the algorithm given above will switch between execution of each replica
at each time
step, $dt$, rather than complete execution of each replica sequentially.
A reordering of the output of the general scheduler, therefore, 
will generally lead to a simpler processor allocation.  Consistent
with the constraints of causality, replica execution in time on a single
processor may be reordered.  Allocation of replicas to processors at each
time step, $dt$, may also be permuted among the processors
as along as the idle time so introduced is tolerable.

Alternatively, the schedule optimization for heterogeneous processors
can be cast as a linear programming problem.  With a penalty for
each switch between replicas on a processor, an optimized schedule may
be derived at the onset by solving the linear programming problem
with a time resolution of $dt$.

\bibliography{scheduler}

\clearpage

\begin{sidewaystable}
\centering
\begin{tabular}{l|c|c|c|c|c|c|c|c|c}
{\footnotesize }&
{\footnotesize $X$}&
{\footnotesize $I$ (\%)}&
{\footnotesize $C$ (\%)}&
{\footnotesize $I$ (\%)}&
{\footnotesize $C$ (\%)}&
{\footnotesize $I$ (\%)}&
{\footnotesize $C$ (\%)}&
{\footnotesize $I$ (\%)}&
{\footnotesize $C$ (\%)}\\
&
&
&
&
{\footnotesize $\gamma=0.1$ }&
{\footnotesize $\gamma=0.1$ }&
{\footnotesize $\gamma=0.5$ }&
{\footnotesize $\gamma=0.5$ }&
{\footnotesize $\gamma=1.0$ }&
{\footnotesize $\gamma=1.0$ }\\
\hline
\hline
{\footnotesize Example 1}\\
{\footnotesize ~~~Maximum efficiency}&
{\footnotesize 12}&
{\footnotesize 0.0}&
{\footnotesize 101.66}&
{\footnotesize $10.55\pm0.02$}&
{\footnotesize $113.81\pm0.05$}&
{\footnotesize $36.96\pm0.06$}&
{\footnotesize $163.26\pm0.23$}&
{\footnotesize $54.55\pm0.10$}&
{\footnotesize $227.59\pm0.46$}\\
{\footnotesize ~~~Minimum run time} &
{\footnotesize 13}&
{\footnotesize 6.16}&
{\footnotesize 100.0}&
{\footnotesize $16.16\pm0.02$}&
{\footnotesize $112.10\pm0.05$}&
{\footnotesize $41.10\pm0.06$}&
{\footnotesize $161.47\pm0.23$}&
{\footnotesize $57.65\pm0.10$}&
{\footnotesize $225.71\pm0.48$}\\
{\footnotesize ~~~Traditional}&
{\footnotesize 20}&
{\footnotesize 39.0}&
{\footnotesize 100.0}&
{\footnotesize $41.12\pm0.03$}&
{\footnotesize $104.12\pm0.06$}&
{\footnotesize $57.35\pm0.06$}&
{\footnotesize $147.18\pm0.29$}&
{\footnotesize $69.65\pm0.07$}&
{\footnotesize $208.72\pm0.60$}\\
{\footnotesize Example 2}\\
{\footnotesize ~~~Maximum efficiency}&
{\footnotesize 3}&
{\footnotesize 0.0}&
{\footnotesize 128.20}&
{\footnotesize $4.65\pm0.02$}&
{\footnotesize $134.62\pm0.07$}&
{\footnotesize $19.29\pm0.09$}&
{\footnotesize $160.33\pm0.33$}&
{\footnotesize $34.50\pm0.73$}&
{\footnotesize $193.76\pm0.66$}\\
{\footnotesize ~~~Minimum run time} &
{\footnotesize 4}&
{\footnotesize 3.93}&
{\footnotesize 100.0}&
{\footnotesize $9.81\pm0.02$}&
{\footnotesize $106.75\pm0.06$}&
{\footnotesize $27.49\pm0.09$}&
{\footnotesize $134.82\pm0.28$}&
{\footnotesize $43.34\pm0.15$}&
{\footnotesize $171.60\pm0.55$}\\
{\footnotesize ~~~Traditional}&
{\footnotesize 20}&
{\footnotesize 80.77}&
{\footnotesize 100.0}&
{\footnotesize $80.62\pm0.01$}&
{\footnotesize $100.00\pm0.10$}&
{\footnotesize $82.55\pm0.02$}&
{\footnotesize $116.60\pm0.31$}&
{\footnotesize $86.41\pm0.03$}&
{\footnotesize $149.97\pm0.48$}\\
{\footnotesize Example 3}\\
{\footnotesize ~~~Maximum efficiency}&
{\footnotesize 6}&
{\footnotesize 0.0}&
{\footnotesize 110.53}&
{\footnotesize $7.25\pm0.02$}&
{\footnotesize $119.28\pm0.05$}&
{\footnotesize $27.70\pm0.07$}&
{\footnotesize $154.55\pm0.22$}&
{\footnotesize $44.20\pm0.13$}&
{\footnotesize $200.60\pm0.44$}\\
{\footnotesize ~~~Minimum run time} &
{\footnotesize 7}&
{\footnotesize 5.26}&
{\footnotesize 100.0}&
{\footnotesize $12.93\pm0.03$}&
{\footnotesize $108.95\pm0.05$}&
{\footnotesize $33.70\pm0.10$}&
{\footnotesize $144.92\pm0.23$}&
{\footnotesize $49.59\pm0.15$}&
{\footnotesize $191.20\pm0.444$}\\
{\footnotesize ~~~Traditional}&
{\footnotesize 50}&
{\footnotesize 86.74}&
{\footnotesize 100.0}&
{\footnotesize $86.75\pm0.01$}&
{\footnotesize $100.78\pm0.11$}&
{\footnotesize $89.07\pm0.03$}&
{\footnotesize $126.69\pm0.38$}&
{\footnotesize $97.76\pm0.03$}&
{\footnotesize $169.56\pm0.66$}\\
\end{tabular}
\caption{Results for the parallel tempering job allocation optimized
by the scheduler
for run time or number of CPUs and for the traditional allocation.
Results are shown for the three 
example systems described in Sec.~\ref{sec:results}.  
Shown are
the number of processors ($X$),  the percentage CPU idle time ($I$),  and
the wall clock time of the simulation relative to the results for 
the traditional allocation without randomness ($C$).
Idle time and wall clock time are also shown for the case where the
CPU time required for each replica is a stochastic quantity,
Eqn.~\ref{eqn:time_random2}, with
$\gamma=0.1, 0.5$, and $1.0$.
\label{table1}}
\end{sidewaystable}

\clearpage

\begin{figure}[htbp]
\caption{Simple example of the allocation of three replicas to
two processors.  In this example, an efficient allocation
requires that replica 2 be split between 
processors 1 and 2.  The replica numbers are marked on the figure.
\label{fig:basic}}
\end{figure}

\begin{figure}[htbp]
\caption{a) Replica allocation in the traditional one replica per processor 
parallel tempering simulation using 20 replicas. 
b) The assignment of the same replicas to 
processors as optimized by the scheduler derived in Sec.~\ref{sec:theory}.  
The replica numbers are marked on the figure.
\label{fig:sched}}
\end{figure}

\begin{figure}[htbp]
\caption{CPU idle time as a function of number of processors used
to solve the 20-replica example 2 from Sec.~\ref{sec:results}.
\label{fig:idle}}
\end{figure}

\newpage

\begin{center}
\epsfig{file=basic.eps,height=3in,clip=,angle=0}
\end{center}
Figure~\ref{fig:basic}.
Earl and Deem, ``Optimal Allocation of Replicas \ldots."

\newpage
\begin{center}
\epsfig{file=sched2.eps,height=3.0in,clip=,angle=0}
\epsfig{file=sched1.eps,height=3.48in,clip=,angle=0}
\end{center}
Figure~\ref{fig:sched}.
Earl and Deem, ``Optimal Allocation of Replicas \ldots."

\newpage
\begin{center}
\epsfig{file=idle.eps,height=3in,clip=,angle=0}
\end{center}
Figure~\ref{fig:idle}.
Earl and Deem, ``Optimal Allocation of Replicas \ldots."

\end{document}